\documentclass{blois}

\def\be{\begin{equation}}
\def\ee{\end{equation}}
\def\bea{\begin{eqnarray}}
\def\eea{\end{eqnarray}}

\newcommand{\Photo}{}

\begin{document}
\vspace*{4cm}
\title{Black hole superradiance 
to search for new particles
}

\author{ D. Blas\footnote{E-mail: \href{dblas@ifae.es}{dblas@ifae.es}}}

\address{{Grup  de  F\'isica  Te\`orica,  Departament  de  F\'isica,\\ Universitat  Aut\`onoma  de  Barcelona,  08193  Bellaterra, Spain}, and \\
{Institut de Fisica d'Altes Energies (IFAE),\\ The Barcelona Institute of Science and Technology,
Campus UAB, 08193 Bellaterra, Spain}}

\maketitle\abstracts{
Rotational superradiance generates the amplification of incoming waves of sufficiently low frequency when scattered with a rotating absorbing body. This may be used to discover new \emph{bosonic} particles of mass $m_b$ if the rotating body has a sufficiently strong gravitational field, that may confine the massive particle and  turn amplification into exponential growth. As a result, the initial seed may be amplified until generating a large cloud around the body, which may have a number of phenomenological consequences.  Rotating black holes are perfect candidates to source this effect, not only from their absorbing and gravitational properties (and hence confining mechanism), but also because for black holes of mass $M_{\rm BH}$, rotational superradiance is efficient for $m_b\sim 10^{-10}\left(\frac{M_{\odot}}{M_{\rm BH}}\right)\rm eV$. The wide range of astrophysical black hole masses brings about new opportunities to probe 
particles of low masses in a large span very hard to detect by any other known method. In this brief contribution I will comment on some of these opportunities.}

\section{Introduction to (black hole) rotational superradiance}

Before introducing rotational superradiance, I want to make clear that this short contribution is by no means a review on this very interesting phenomenon. For this, I strongly recommend\cite{Bekenstein:1998nt,Brito:2015oca} together with the original references \cite{Zel,Teukolsky:1974yv}. Furthermore, the constraints that apply to these proceedings also mean that I can not fairly represent all the interesting works in this field of research in recent years. My hope is to trigger the curiosity of the reader to explore more by herself/himself \footnote{I recommend the reader to have a look at the recent presentations by two of the leaders in the field
\url{https://www.tat.physik.uni-tuebingen.de/tatseminar/Slides/20211210 Brito.pdf} and \url{https://indico.kias.re.kr/event/89/}.
}. The Planck mass is set to 1 in this note. Also $\hbar=c=1$.

After this preamble, we can follow\cite{Zel} and introduce \emph{rotational superradiance} (\emph{superradiance} from now on \footnote{The term `superradiance' was coined by R. Dicke in \cite{Dicke:1954zz} in a different context, and is a standard term in quantum optics, see also \cite{Bekenstein:1998nt}. Since no confusion will arise in this article, I will follow the commonly accepted terminology of calling rotational superradiance  simply `superradiance'.}): given a wave propagating radially and with azimuthal number $m$,
\be
\Phi \sim A_i e^{-i \omega(t+r)} e^{i m \phi} S(\theta),
\ee
when scattered from an absorbing rotating body of rotational frequency $\Omega$ satisfying
\be
\omega < m\Omega, \label{eq:SRc}
\ee
the wave is \emph{amplified} (technically, the reflection coefficient is larger than one). The energy for this amplification is extracted from the rotation of the absorbing body.

 Three extra ingredients are necessary to make this process \emph{more} interesting for fundamental physics: $i)$ adding a mass $m_b$ to the incoming wave (this, together with point $iii)$ will confine the wave to the gravitational object and generate an exponential growth) $ii)$ making it correspond to a \emph{bosonic} degree of freedom (to allow for large occupation numbers)\footnote{For the description of superradiance for fermions and its connection to the Klein paradox see  \cite{Brito:2015oca}.} and $iii)$ considering a rotating body with strong dissipation and gravitational field. In this case, superradiance can be  tremendously amplified by generating what is known as a \emph{cloud} of the incoming field with huge occupation numbers. Point $iii)$ is satisfied automatically by rotating\footnote{Technically, the presence of an ergoregion plays an important role for BH superradiance\cite{Vicente:2018mxl}.} black holes (BHs), which is one of the reasons why I will only discuss `black hole superradiance' (see e.g. \cite{Cardoso:2017kgn,Chadha-Day:2022inf} for a related phenomenon in stars). Several details about this process and the final configuration can be found in the review\cite{Brito:2015oca}. Here I will only give some simple estimates for concrete situations.

 One of the most important aspects of black hole superradiance  in what respects the growth of a cloud, is that it is efficient  when 
 \be
 m_b M_{\rm BH} \sim \frac{m_b}{10^{-10} \mathrm{eV}} \frac{M_{\rm BH}}{M_{\odot}} \sim O(1), \label{eq:O1}
 \ee
 where $M_{\rm BH}$ is the mass of the BH. This is a manifestation of the process being related to the appearance of a bound state for particles of mass $m_b$ in the gravitational potential of the black hole. For instance, the time of generation of the cloud for $ m_b M_{\rm BH} \ll 1$  in the case of a scalar\footnote{One can also consider  new massive particles of spin-1 or spin-2. The spin degree of freedom changes parametrically the predictions, though the main picture remains universal.} has been found to be $\tau \approx 30~ {\rm days }\left(\frac{M_{\rm BH}}{10 M_{\odot}}\right)\left(\frac{0.1}{M  m_b}\right)^9\left(\frac{0.9}{\chi_i}\right)$, where $\chi_{i}$ refers to the initial spin of the black hole,  see e.g. \cite{Baumann:2019eav}. The parametric form of this equation is not important for the present discussion, just the fact that it is very short\footnote{ This has lead to the claim that any small amount of the massive field (even a quantum fluctuation) would easily generate the cloud. } in astrophysical timescales only  if the condition (\ref{eq:O1}) is not badly broken. The rest of the estimates in this section will be performed in the limit where $m_b M_{\rm BH}\ll 1$ (though not extremely small, since otherwise $\tau$ becomes too large). 
  The radius of the cloud in this case is given by $r\sim 1/(M_{\rm BH} m_b^2)$ (another manifestation of this being a non-relativistic bound state).   
  Also, a significant fraction of the total energy of the black hole may be converted into the cloud, as much as $0.1 M_{\rm BH}$. The previous two points imply the large energy density, 
 \begin{equation}
 \rho_b\sim 0.1 M_{\rm BH}/r\sim 10^{48} \frac{M_{\rm BH}}{M_{\odot}}\left(\frac{m_b}{10^{-10} \mathrm{eV}}\right)^3 \mathrm{eV} / \mathrm{cm}^3. \label{eq:density}
 \end{equation}
The cloud has orbital quantum numbers  $l\geq 1$, with $l=1$ typically dominant. These two last aspects are important for the emission of gravitational waves. In the following sections, I will focus on some possible ways to detect the cloud (see also \cite{Arvanitaki:2010sy} for early ideas in this direction).

 \section{Phenomenological consequences  of superradiance}

From the previous section, since the cloud of bosons may be as massive as  $0.1 M_{\rm BH}$, its influence in the dynamics of black holes of mass $M_{\rm BH}$ may be significant. As a consequence, the rotation of the latter may be affected. Furthermore, any orbital motion around the black hole should also be modified by the existence of this very asymmetric cloud. Finally, the dynamics of the cloud is also very interesting: as it rotates,  it generates a large time-dependent quadrupole sourcing gravitational waves; it can also decay among different modes either emitting gravitational waves or other particles that may be coupled to the new boson. 

For this fantastic phenomenology to be possible  with astrophysical black holes (in particular with masses $10^9 \, M_{\odot}>M_{\rm BH}>M_{\odot}$), one sees from   (\ref{eq:O1}) that the new bosons should be of very low mass \footnote{One can also consider the consequences for e.g. pions or even the Higgs particle if very small black holes are somehow generated in the early Universe, but I won't discuss this possibility here. See, e.g. \cite{Ferraz:2020zgi}.}. If this corresponds to the \emph{fundamental mass}, then one necessarily has to consider particles beyond the standard model. This is hardly a problem, since there are very well motivated ultra-light particles appearing in physics beyond the standard model (see e.g. \cite{Hui:2016ltb}), which are \emph{very hard} to detect by any other method\footnote{For instance, these particles naturally appear with very small couplings to the standard model, and as dark matter candidates carry very small momentum.}. Furthermore, an ``effective'' mass (a gap in the dispersion relation) can also be generated by the propagation of the boson in a medium (recall that the purpose of the mass is to make the wave `fall back' into the black hole and generate bound states). As I will discuss briefly, the mass of low frequency photons in the intergalactic medium has a bulk value at cosmological times within the range to generate superradiant clouds around astrophysical BHs.  Let me discuss these two cases a bit more in detail.

 \subsection{Looking for superradiance of fundamental fields: gravitational effects}

The first searches I want to discuss are those related to the \emph{Regge plane}, namely the distribution of spins for black holes of different masses. The logic is relatively straightforward: when a cloud is formed, it  reduces the angular momentum of the back hole until the condition (\ref{eq:SRc}) is no longer satisfied. As a consequence, given a new particle with mass $m_b$, the black holes satisfying (\ref{eq:O1}) should not be spinning very fast. This is true unless another mechanism (for instance, accretion) spins them up, which complicates the analysis. Still, the time scales of the spin-down related to superradiance are so short that the observation of any highly spinning black hole should be a rather convincing evidence to exclude certain masses. As a result, once the Regge plane is populated by observations of spins of astrophysical black holes (in particular from the observations of gravitational waves), one may be able to constrain the presence of new bosons of masses in the range $(10^{-12}\rm eV, 10^{-19}\,\rm eV)$ \cite{Brito:2017zvb}. A precise analysis in this direction from LIGO-Virgo GWTC-2  data can be found in\cite{Ng:2020ruv} (as the reader may have guessed, it corresponds to masses around $10^{-13}$\,eV).

Another rather direct consequence of the existence of the superradiant cloud is the emission of gravitational waves through different channels\footnote{Another interesting signal of GWs created by superradiance is related to the (very strong) emission of GWs by black hole mimickers \cite{Barausse:2018vdb}.}. The most straightforward and universal signal comes from the time dependent quadrupole of the cloud itself. The emitted gravitational waves have characteristic frequency $\omega_{\rm gw}=2 m_b$, with characteristic decay time (for the case of scalar bosons)
\be
\tau_{\mathrm{GW}}^S \approx 10^5 \mathrm{yr}\left(\frac{M_{\rm BH}}{10 M_{\odot}}\right)\left(\frac{0.1}{M_{\rm BH} m_b}\right)^{15}\left(\frac{0.5}{\chi_i-\chi_f}\right),
\ee
where $\chi_{i/f}$ refers to the initial/final spin of the black hole. These estimates are discussed in detail in e.g.\cite{Tsukada:2020lgt}. A big challenge in modelling this signal comes from the need to understand the population of BHs of given masses. This has large astrophysical uncertainties. Still, quite remarkably, future interferometers may detect these signals if new bosons with masses in the appropriate range do exist (see also \cite{Brito:2017wnc} for analysis of the stochastic signal expected for scalar fields).  Another possible signal from the clouds arises from energy emitted from the transitions between excited levels of the cloud itself \cite{Arvanitaki:2010sy}. Their possible detection has also triggered growing interest in the community, and may offer a unique handle in the existence of new particles of masses $10^{-17}-10^{-12}$\, eV
\cite{Zhu:2020tht,Juliana:2022fbd}. Finally, one may wonder about what happens if the BH is not isolated, but is part of a binary system. In this case, the phenomenology is even richer, with possible new transitions, the possible depletion of the cloud, the modification of the binary dynamics from the existence of the cloud and other new phenomena, see eg. \cite{Baumann:2022pkl,Berti:2019wnn,Hannuksela:2018izj}. 

Altogether, there are realistic chances of detecting the superradiant clouds associated to the existing of new particles or forbidding the existence of these particles in a wide range of masses. Thinking that we can comprehensibly explore the presence of new particles in this way is mind-blowing. These prospects have generated a significant amount of activity in the field summarized in the references before (see also \cite{KAGRA:2021tse} and a battery of references in my slides in \url{https://indico.cern.ch/event/1133536/contributions/4834373/}).

 \subsection{Looking for superradiance of fundamental fields: other astrophysical  effects}
 
 The universality of gravitation means that the conclusions of the previous section are very generic.  Furthermore, when a new particle is motivated for some reason (e.g as a candidate of dark matter, related to the strong CP problem, origin of neutrino masses, etc) it is expected that it will also interact with  fields of the standard model (photons, gluons, quarks, leptons...) or itself. This opens new directions to use superradiance for fundamental physics. The self-interactions of the boson may generate extra decays of the cloud, or even explosive (bose-nova) events \cite{Arvanitaki:2010sy}. In this section I will only discuss the situation where the new boson $a$ is a pseudo-scalar particle, and a coupling with electromagnetism of the form
 \be
 g_a a \tilde{F}_{\mu \nu} F^{\mu \nu},
 \ee
 is expected. In this situation, it has been shown that for a large range of values of $g_a$ (basically for 
 $g_a > 10^{-19} \sqrt{\frac{M_{\rm BH}}{M_a}} \frac{1}{\left(m_a M_{\rm BH}\right)^2}\, \mathrm{GeV}^{-1}$, $M_a$ being the  mass of the cloud), the cloud may explosively decay into pairs of photons \cite{Ikeda:2018nhb,Rosa:2017ury}. The fate of these photons was studied in \cite{Blas:2020nbs}. The picture is the following:  if the medium surrounding the BH contains free charged particles, these very low frequency photons will be absorbed almost instantaneously by inverse Brehmstrahlung. This absorption will inject the tremendous energy 
 \be
 \frac{d E}{d t} \sim 10^{66}\left(\frac{M_a}{M_{\rm BH}}\right) \mathrm{eV} / \mathrm{s},
 \ee
 into the surrounding plasma, heating it, and generating a burst of ultra-hot baryons/electrons. As discussed in  \cite{Blas:2020nbs}, for this to happen one requires the mean free path to be short enough, and the mass of the axion to be above the plasma mass. For the astrophysical phenomenology of interest, this reduces the range of masses of interest to $m_a\sim(10^{-13}-10^{-10})$\,eV. Our estimates show that this burst generates a bubble of up to Mpc size at redshifts before or after reionization. As a result, this bubble may not only emit high energy radiation, but will also contribute to the reionization of the Universe, and generate spectral distortions in the CMB. Our conclusion of  \cite{Blas:2020nbs} is that a \emph{single} bubble may have visible consequence for the next generation of CMB observations, and, even better, if several bubbles are produced their detection is almost guaranteed. Still, the difficulty lays in understanding the population of BHs at high redshift, something we would like to include in a future analysis.
 
 Before moving to the next section, I would like to mention the less known fact that black hole superradiance may also impact the searches for dark photons through their interaction with light, see \cite{Caputo:2021efm}.

 \subsection{Superradiance of photons?}
 
The superradiance phenomenon, in particular the `bomb' effect of massive fields in the gravitational field of a black hole, requires the wave to correspond to a field that can be bounded to the gravitational source. This is not possible for massless fields, unless the dispersion relation is modified by the medium. As it is well known, the simplest case where a \emph{photon} may be considered massive is when it propagates in a ionized medium with  frequencies above the plasma mass \footnote{In principle, more complex dispersion relations may also generate a cloud, something which to my knowledge has not been systematically explored. I also want to remark that there has been some work showing that the plasma mass is not completely equivalent to a Proca mass term, which may have some implications for superradiance \cite{Cannizzaro:2020uap}.}. Quite remarkably (almost `miraculously'), when one considers the plasma mass induced by the intergalactic medium, its value 
\be
m_\gamma \simeq \omega_p=\sqrt{\frac{4 \pi \alpha n_e}{m_e}} \sim 10^{-10} \sqrt{\frac{n_e}{\mathrm{~cm}^3}} \mathrm{eV},
\ee
for the expected average number density of free electrons $n_e(z)$ from $z\approx10^4$ to $z\sim 0$, corresponds to BHs of astrophysical mass after using (\ref{eq:O1}), \cite{Pani:2013hpa,Conlon:2017hhi,Blas:2020kaa}. One can then dream of the possibility that astrophysical black holes could be used as \emph{probes} of the intergalactic medium, or that this phenomenon could be a way to understand the population of astrophysical black holes!

Unfortunately, this possibility does not seem to be realized in nature. Some early criticism was related to  the fact that the effective mass in an accreting situation would not be homogeneous, though the work \cite{Dima:2020rzg} showed that this was not necessarily a problem. More relevant is  the fact that several mechanisms are at play when a very dense cloud of photons (cf. 
(\ref{eq:density})) is generated, which eventually quenches superradiance. This was  first realized in \cite{Fukuda:2019ewf}, where they considered the dissipation of the cloud through  pair creation, and latter in \cite{Cardoso:2020nst,Blas:2020kaa} where it was shown that non-linear effects and other electrodynamical effects (e.g the acceleration of the surrounding electrons) quench the cloud at times many orders of magnitud faster than the typical time scale required for the cloud to be generated. 

 \section{Conclusions}

Rotational superradiance offers an extraordinary handle into physics beyond the standard model. It complements other searches of new particles by accessing the (otherwise very hard to constrain) mass range of ultralight particles with masses $m_b\in 10^{-18}-10^{-12}$\,eV. Furthermore, the phenomenon is extraordinary in its almost independence of the initial density of the new field. As a result, one can not only explore dark matter models (where the energy density of the field is already present), but \emph{any} candidate with the correct mass. 

The phenomenology is extremely rich, and one expects a collection of disparate signals, as   coherent radiation, stochastic gravitational wave backgrounds, modifications of the Regge plane, large explosions able to heat up bubbles of Mpc size (a manifestation of the large amount of energy in the cloud)... In my opinion, there are still many new possibilities to explore, in particular if the new boson interacts with the fields of the standard model. Still, the most spectacular progress will come once LISA flies and the next generation of Earth-based gravitational wave observatories start to work. This is because despite the uncertainties in the populations of BHs at different cosmological times, most models predict that future data will be sensitive to some of the implications of superradiant clouds. 

Several puzzling aspects of our universe (dark matter, strong CP-problem, string theory landscape...) points towards the existence of new ultralight fields. Superradiance will play a unique role in the search for these new building-blocks of Nature, and it is even possible that it will be \emph{the} channel to discover them.

\section*{Acknowledgments}

 It is a pleasure to thank A. Caputo, P. Pani and S. Witte for all their great work, generosity and inspiration in our works together. I am also grateful to R. Brito,  E. Barausse and R. Vicente for their comments on this draft. Finally, I am grateful to R. Brito, V. Cardoso and P. Pani for their effort to write the impressive reference\cite{Brito:2015oca} that made the life of many researchers much easier when approaching black hole superradiance. 
 IFAE is partially funded by the CERCA program of the Generalitat de Catalunya.  DB is supported by a `Ayuda Beatriz Galindo Senior' from the Spanish `Ministerio de Universidades', grant BG20/00228. 
The research leading to these results has received funding from the Spanish Ministry of Science and Innovation (PID2020-115845GB-I00/AEI/10.13039/501100011033).

%\section*{Appendix}
%
% We can insert an appendix here and place equations so that they are
%given numbers such as Eq.~\ref{eq:app}.
%\be
%x = y.
%\label{eq:app}
%\ee

\end{document}